\pdfoutput=1

\documentclass[11pt]{article}

\usepackage[final]{acl}

\usepackage{times}
\usepackage{latexsym}

\usepackage[T1]{fontenc}

\usepackage[utf8]{inputenc}

\usepackage{microtype}

\usepackage{inconsolata}

\usepackage{graphicx}

\usepackage{threeparttable}
\usepackage{amsmath}
\usepackage{amsfonts}
\usepackage{booktabs}

\title{Enhancing Polyglot Voices by Leveraging Cross-Lingual Fine-Tuning in Any-to-One Voice Conversion}

\author{
  \textbf{Giuseppe Ruggiero\textsuperscript{1,2}},
  \textbf{Matteo Testa\textsuperscript{2}},
  \textbf{Jurgen Van de Walle\textsuperscript{2}},
  \textbf{Luigi Di Caro\textsuperscript{1}}
\\
  \textsuperscript{1}Università degli studi di Torino, Turin, Italy \\
  \textsuperscript{2}Cerence Inc, Turin Italy
\\
    \{giuseppe.ruggiero, luigi.dicaro\}@unito.it \\ \{matteo.testa, jurgen.vandewalle\}@cerence.com
}

\begin{document}
\maketitle
\begin{abstract}
The creation of artificial polyglot voices remains a challenging task, despite considerable progress in recent years. This paper investigates self-supervised learning for voice conversion to create native-sounding polyglot voices. We introduce a novel cross-lingual any-to-one voice conversion system that is able to preserve the source accent without the need for multilingual data from the target speaker. In addition, we show a novel cross-lingual fine-tuning strategy that further improves the accent and reduces the training data requirements. Objective and subjective evaluations with English, Spanish, French and Mandarin Chinese confirm that our approach improves on state-of-the-art methods, enhancing the speech intelligibility and overall quality of the converted speech, especially in cross-lingual scenarios.\footnote{Audio samples are available at: \url{https://giuseppe-ruggiero.github.io/a2o-vc-demo/}}
\end{abstract}

\section{Introduction}
\label{sec:introduction}
 \begin{figure*}[t]
  \centering
  \includegraphics[width=0.95\linewidth]{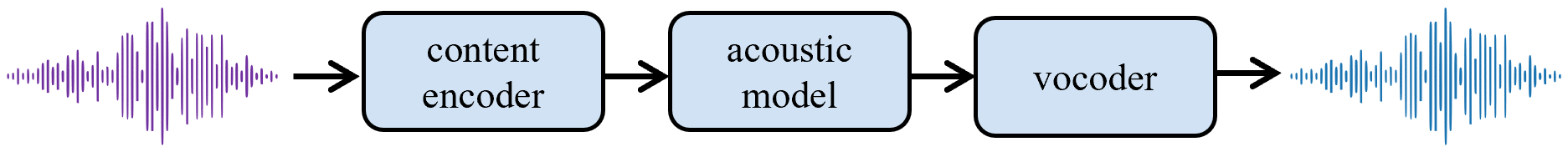}
  \caption{High-level overview of the voice conversion system.}
  \label{fig:architecture}
\end{figure*}
In recent years, advancements in speech synthesis~\citep{vallex, openvoice, softvc} have significantly improved the quality of artificial polyglot voices. Several noteworthy studies in cross-lingual voice conversion (XVC) merit attention, especially some any-to-one (A2O VC)\footnote{The aim of A2O VC is to convert speech from any speaker, including those unseen during training, to that of a fixed target speaker.} variants~\citep{softvc, s3prl-vc}. The aim of XVC is to convert source voices into target voices without changing the linguistic content, even when source and target speak different languages~\citep{xvc}. From an information perspective, VC can be accomplished by initially disentangling the source speech into representations describing the linguistic content and speaker characteristics and then synthesizing the converted speech from the extracted linguistic content while incorporating the identity of the target speaker~\citep{s3prl-vc}.

To disentangle content and speaker information, an ASR model trained on a labeled dataset is frequently employed to extract the supervised spoken content representation, such as text~\citep{cascading-asr-tts} or phonetic posteriorgram (PPG)~\citep{ppg}. However, despite its promising results in VC, this approach encounters several limitations. First, the assembly of labeled datasets can be prohibitively expensive, especially in low-resource settings like the XVC scenario~\citep{vcc2020}. Second, the system's sensitivity to ASR errors can lead to potential mispronunciations or inaccuracies in conversion. Finally, text and PPG do not capture all linguistic nuances such as accent, expressions and style resulting in neutral-sounding synthesized speech~\citep{ace-vc}.

In several recent works~\citep{softvc, s3prl-vc, speech-resynthesis}, self-supervised learning (SSL) is used to train models for speech synthesis. This approach is primarily driven by the following advantages offered by SSL: 1) It requires only large-scale unlabeled data to learn rich and compact speech representations; 2) Numerous large-scale pre-trained SSL models are publicly available, including Wav2vec2~\citep{wav2vec2}, HuBERT~\citep{hubert}, and WavLM~\citep{wavlm}; 3) These models have been shown to generalize well across various languages, even though some of them have been trained on English data only. Since SSL representations contain a significant amount of information, including phonetic and speaker details, many studies~\citep{s3prl-vc, speech-resynthesis} have attempted to discretize them to extract disentangled linguistic content. However, while effective at removing speaker information, discretizing SSL representations results in lossy compression, which leads to mispronunciations in the converted speech~\citep{contentvec}.

To address this issue, alternative disentanglement strategies have been proposed. Soft-VC \citep{softvc} creates soft speech units starting from discrete ones, which are obtained by applying k-means~\citep{kmeans} clustering over the output of a pre-trained HuBERT model. Specifically, a network is trained to predict a distribution over the discrete units to preserve more content information and thus correct mispronunciations. Neural Analysis and Synthesis (NANSY)~\citep{nansy} introduces information perturbation (prior to the SSL model) as a way to disentangle speaker-related information from other aspects of speech. In \citet{per-utt-std}, utterance-level standardization is used to remove speaker-specific information. However, while these methods are better at improving intelligibility than discretization, our findings indicate that the linguistic content is still compromised, especially in cross-lingual scenarios. Furthermore, other speaker-related features (e.g., the accent) are also distorted, as highlighted in \citet{contentvec}. 

Recently, a variety of more versatile systems have emerged. One such example is kNN-VC \citep{knn-vc}, an any-to-any VC (A2A VC) approach that extracts frame-wise WavLM representations for both source and target speech. It then converts by directly replacing each frame of the source representation with its nearest neighbors in the target representation set. This strategy allows SSL features to be used without any further disentanglement step.

We propose an A2O VC system similar to that of~\citet{softvc} except it learns to disentangle content information without the need for supplementary steps like discretization, and it can both enhance cross-lingual generation and reduce the training data requisites. Our main contributions are as follows: 
\begin{itemize}
    \item We show that a bottleneck information extractor similar to that of~\citet{freevc} is sufficient to improve intelligibility in cross-lingual scenarios without compromising on speech quality and target speaker identity.
    \item We propose a cross-lingual pre-training fine-tuning strategy. This approach not only further enhances VC performance in cross-lingual scenarios but also allows low-quality data to be used for data intensive pre-training.
 \end{itemize}

\section{Model Architecture}
\label{sec:model-architecture}
The system consists of three components: a content encoder, an acoustic model, and a vocoder (Figure~\ref{fig:architecture}). The content encoder extracts speech representations from an audio of any speaker, the acoustic model translates these representations into a mel spectrogram of the target speaker, and the vocoder converts the resulting mel spectogram into a time-domain waveform. For both training and inference, only audio data is required.

\subsection{Problem Definition}
\label{subsec:problem-definition}
Consider a dataset \textit{D} of a target speaker containing \textit{M} utterances in the time domain. Let us denote the $i$-th utterance of the target speaker as $u_i$, its features extracted by a pre-trained SSL model as $v_i \in \mathbb{R}^s$, and its mel spectrogram as $x_i$, where \(i \in [1, M]\). 
The content encoder $\mathcal{C}$ takes $u_i$ as its input and generates $v_i$:
\begin{align}
    v_i=\mathcal{C}( u_i ; w_{\mathcal{C}} )
    \label{eq:content-encoder-eq}
\end{align}
where $w_{\mathcal{C}}$ represents the content encoder parameters. To remove speaker information in $v_i$ while minimizing its impact on other components (e.g., linguistic content), the acoustic model $\mathcal{A}$ uses a pre-net $\mathcal{P}$ which projects $v_i$ into a space $\mathbb{R}^d$, where $d \ll s$. This dimensionality gap imposes an information bottleneck, which allows the acoustic model to discard content-irrelevant information such as noise and speaker identity~\citep{freevc}. Thus, the acoustic model $\mathcal{A}$ predicts $\hat{x}_i$:
\begin{align}
    \hat{x}_i=\mathcal{A}( \mathcal{P}( v_i; w_{\mathcal{P}} ) ; w_{\mathcal{A}} )
    \label{eq:acustic-model-eq}
\end{align}
where $w_{\mathcal{P}}$ represents the parameters of the pre-net, while $w_{\mathcal{A}}$ represents those of the acoustic model. Finally, the vocoder $\mathcal{V}$, given $\hat{x}_i$, generates $\hat{u}_i$:
\begin{align}
    \hat{u}_i=\mathcal{V}( \hat{x}_i ; w_{\mathcal{V}} )
    \label{eq:vocoder-eq}
\end{align}
where $w_{\mathcal{V}}$ represents the vocoder parameters. 

The three components of the VC system are trained independently. The content encoder $\mathcal{C}$ is an off-the-shelf model pre-trained on a large set of multi-speaker data whereas the acoustic model $\mathcal{A}$ and the vocoder $\mathcal{V}$ are trained only on target speaker data. After generating pairs $(v_i, x_i)$ for each $u_i$ of the dataset \textit{D}, the acoustic model $\mathcal{A}$ is trained by optimizing the following objective function:
\begin{align}
    \min _{w_{\mathcal{P}}, w_{\mathcal{A}}} L_{\mathcal{A}}( x_i, \mathcal{A}( \mathcal{P}( v_i; w_{\mathcal{P}} ) ; w_{\mathcal{A}} ) ))
    \label{eq:acoustic-model-loss}
\end{align}
where $L_{\mathcal{A}}$ is a loss function in the time-frequency domain such as the $L_1$ distance between target and predicted mel spectrograms. 

The vocoder $\mathcal{V}$ is trained either directly on ground truth mel spectrograms or on the mel spectrograms predicted by the acoustic model:
\begin{align}
    \min _{w_{\mathcal{V}}} L_{\mathcal{V}}( u_i, \mathcal{V}( x_i ; w_{\mathcal{V}} ) ) \text { or } \min _{w_{\mathcal{V}}} L_{\mathcal{V}}( u_i, \mathcal{V}( \hat{x}_i ; w_{\mathcal{V}} ) )
    \label{eq:vocoder-loss}
\end{align}
where $L_{\mathcal{V}}$ is a loss function in the time domain. 

By training only on target speaker data, the loss function in Eq.~\ref{eq:acoustic-model-loss} makes the acoustic model target speaker specific: it generates spectrograms of that speaker. The primary objective is to convert the source speaker's identity into that of the target speaker. Consequently, it is possible to perform XVC by converting a source utterance from a language different from that spoken by the target speaker, thereby enabling the target speaker to become polyglot.


\subsection{Cross-Lingual Fine-Tuning}
\label{subsec:cross-lingual-fine-tuning}
Although the system achieves intra- and cross-lingual A2O VC with good intelligibility and high quality overall, some phonemes are mispronounced or colored by the accent of the target speaker, especially in cross-lingual scenarios. To tackle this issue, we opted for a hybrid approach, leveraging the phonetic knowledge of a model trained on a specific language (e.g., French) to enhance the pronunciation and linguistic content of the target speaker (e.g., an English speaker) when it speaks that language. Let us consider a \textit{lang1}-\textit{lang2} cross-lingual scenario. Our objective is to enhance the acoustic model on \textit{lang1} without compromising the intra-lingual and other cross-lingual tasks.

In addition to the target speaker dataset $\textit{D}$ ($\textit{lang2}$), let us also consider a dataset $\textit{D'}$ consisting of $\textit{N}$ speakers, each of whom has \textit{M'} utterances in $\textit{lang1}$. Let us denote the $i$-th utterance of the $j$-th speaker as $u'_{ji}$, its SSL vector as $v'_{ji}$, and its mel spectrogram as $x'_{ji}$, where \(j \in [1, N]\) and \(i \in [1, M']\). 

First, we pre-train the acoustic model ${\mathcal{A}}$ using the loss in Eq.~\ref{eq:acoustic-model-loss} but employing pairs $(v'_{ji}, x'_{ji})$ for each $u'_{ji}$ from the dataset $\textit{D'}$. Eq.~\ref{eq:acoustic-model-loss} becomes:
\begin{align}
    \min _{w'_{\mathcal{P}}, w'_{\mathcal{A}}} L_{\mathcal{A}}( x'_{ji}, \mathcal{A}( \mathcal{P}( v'_{ji}; w'_{\mathcal{P}} ) ; w'_{\mathcal{A}} ) ))
    \label{eq:acoustic-model-loss2}
\end{align}
Second, we fine-tune ${\mathcal{A}}$ using the dataset $\textit{D}$ (or a subset thereof) from the target speaker, as explained in Section \ref{subsec:problem-definition}. This allows the acoustic model to acquire phonetic coverage of $\textit{lang1}$ during the pre-training, which can be leveraged during the fine-tuning on $\textit{lang2}$. Consequently, the $\textit{lang2}$ target speaker will exhibit improved $\textit{lang1}$ linguistic characteristics after the conversion.

The cross-lingual fine-tuning technique offers several advantages. It enhances the linguistic content and pronunciation in XVC and reduces the amount of training data needed from the target speaker. While training the acoustic model without fine-tuning typically requires at least 10-12h of high-quality audio from the target speaker, the fine-tuning only requires 2h. 

\subsection{Content encoder}
\label{subsec:content-encoder}
The content encoder transforms an input waveform into a more concise and information-dense representation known as SSL vector. We chose WavLM as pretrained SSL model. WavLM was pre-trained by Microsoft using a self-supervised strategy similar to that of \citet{hubert} on 60,000 hours of Libri-Light \citep{libri-light}, 10,000 hours of GigaSpeech \citep{gigaspeech} and 24,000 hours of VoxPopuli \citep{vox-populi}, for a total of 94,000 hours of audio data. We used a publicly available WavLM-Large\footnote{\url{https://huggingface.co/microsoft/wavlm-large}} model employing the output of its 15th transformer layer as the representation for speech audio $v_i$ in all of our experiments. This decision was motivated by the findings of \citet{how-can-a-bad-teacher, wav2vec-u}, which demonstrated the efficacy of this layer in phone discrimination tests. Accordingly, \textit{s} in Subsection \ref{subsec:problem-definition} is configured to 1024, that is the dimension of the output vectors of WavLM-Large.

\subsection{Acoustic Model and Vocoder}
\label{subsec:acoustic-model-and-vocoder}
The acoustic model and vocoder are typical components in a text-to-speech (TTS) system. 

The proposed acoustic model takes as input SSL vectors rather than graphemes or phonemes, and outputs target speaker mel spectrograms. Following~\citet{softvc}, the model's architecture resembles that of Tacotron 2~\citep{tacotron2} as it features an encoder and an autoregressive decoder. Both the encoder and decoder are preceded by a feed-forward pre-net, while a final linear layer with $n$-MELs units follows the decoder. No attention module is used~\citep{s3prl-vc}. The encoder pre-net is a feed-forward neural net composed of a stack of two linear layers with 256 units, ReLU activations, and dropout~\citep{dropout}. The encoder consists of a stack of three 1D-convolutional layers with 512 units, kernel size 5, stride 1, padding 2, and ReLU activations. Following~\citet{adainvc}, we added an Instance Normalization (IN) layer after each Conv1D layer, as this helps to further reduce speaker identity while preserving content information. The decoder predicts each spectrogram frame from the output of the encoder and the previously generated frames. First, a decoder pre-net, similar to the encoder one, is applied and then three LSTMs with 768 units each followed by a final linear layer with $n$-MELs units are used.

The vocoder transforms the predicted mel spectrograms into audio waveforms. We used a HiFi-GAN~\citep{hifigan} model, a state-of-the-art real-time neural vocoder. 

\subsection{Length Regulator}
\label{subsec:length-regulator}
As there is no attention mechanism, we use an optional duration adjustment strategy to mitigate potential length mismatches between the SSL input features and the target spectrogram sequence. Indeed, WavLM generates features at a 20 ms frame rate for audio sampled at 16 kHz (resulting in a hop length of 320), which might lead to a length mismatch when generating spectrograms with different time resolutions (e.g., different hop lengths and/or higher sample rates). This strategy involves integrating a length regulator module between the encoder and the decoder of the acoustic model which interpolates the SSL features extracted from the source speech. To do this, we leverage PyTorch's \textit{interpolate}\footnote{\url{https://pytorch.org/docs/stable/generated/torch.nn.functional.interpolate.html}} function with its default settings. 

\section{Experiments}
\label{sec:experiments}
To evaluate the effectiveness of our VC approach, experiments across a range of different tasks and datasets were conducted. All the experiments were run on a Linux machine with a single NVIDIA Titan RTX GPU with 24 GB of RAM.

The proposed system was tested as two variants: \textit{Proposed}, our VC method using the conventional training strategy, and \textit{Proposed-F}, the alternative system employing the cross-lingual fine-tuning strategy. We selected two baseline models for comparison: Soft-VC\footnote{\url{https://github.com/bshall/soft-vc}}~\citep{softvc} and kNN-VC\footnote{\url{https://github.com/bshall/knn-vc}}~\citep{knn-vc}. Both excel in cross-lingual voice conversion tasks, demonstrating strong content intelligibility, accent preservation, and speech quality.

\subsection{Experimental Setup}
\label{subsec:experimental-setup}
We conducted experiments for both intra- and cross-lingual settings using LJSpeech~\citep{ljspeech} as the target speaker dataset. We selected four languages for our evaluation: American English, European French, European Spanish, and Mandarin Chinese. For English, we used LibriSpeech~\citep{librispeech}; for French and Spanish, we used Multilingual LibriSpeech (MLS)~\citep{mls}; for Mandarin Chinese, we used Aishell~\citep{aishell}. 

To ensure a fair comparison, when training the LJSpeech model we used the same configuration as~\citet{softvc}, including their HiFi-GAN vocoder pre-trained on LJSpeech, unless otherwise stated. More specifically, we used a 16 kHz sampling rate and we set $n$-MELs to 128, $s$ to 1024 and $d$ to 256. 

For cross-lingual fine-tuning, we initially pre-trained three distinct acoustic models for each language using randomly selected 150h of audio data from the training sets of LibriSpeech, French MLS, Spanish MLS, and Aishell, respectively. The choice of 150h was based on Aishell's training set size in order to ensure that each language gets the same amount of training hours. Each pre-trained acoustic model was fine-tuned with only 2h of LJSpeech data for 15k steps using a batch size of 8. We found this number of hours to be optimal for the fine-tuning step through several experiments conducted using a validation set composed of 100 utterances from LJSpeech. We also observed that using less than 2h of data was insufficient to obtain a good overall quality.

\subsection{Evaluation Metrics}
\label{subsec:evaluation-metrics}
We assessed intelligibility, speaker similarity, accent preservation and overall speech quality in subjective and objective evaluations. 

We created a test set for each language to be used as the source speech for conversion into LJSpeech. Each set consists of 60 utterances obtained by randomly selecting 3 utterances from 20 speakers (10 male and 10 female) extracted from: the test-clean set of LibriSpeech for English; the test set of Spanish MLS for Spanish; a set composed of the dev and test sets of French MLS combined for French; and a set composed of the dev and test sets of Aishell combined for Chinese.

\subsubsection{Objective Evaluations}
\label{subsubsec:objective-evaluations} 
To evaluate the \textit{intelligibility} of the converted speech, we measured word error rate (WER) and phoneme error rate (PER) between the source and converted speech using the Whisper Medium ASR model\footnote{\url{https://huggingface.co/openai/whisper-medium}}~\citep{whisper} for English, French and Spanish and the FunASR model\footnote{\url{https://github.com/modelscope/FunASR}}~\citep{funasr} for Chinese. Lower error rates indicate more intelligible speech, as the original words/phonemes are still recognizable after conversion. We measured \textit{speaker similarity} (SSIM) by using a trained speaker verification model\footnote{\url{https://github.com/resemble-ai/Resemblyzer}}. Specifically, we computed the cosine similarity between the d-vectors~\citep{d-vector} of each converted sample and that of the target speaker. The target speaker embedding is generated by averaging the d-vectors of 50 utterances from LJSpeech.

\subsubsection{Subjective Evaluations}
\label{subsubsec:subjective-evaluations} 
We conducted subjective evaluations to assess both the \textit{overall quality} and the \textit{native accent level} of the converted audio. 

For each language, we selected all 60 utterances from the corresponding test set and converted them into LJSpeech with our two proposed methods and the two baselines, resulting in a total of 300 utterances per language (60 ground truth + 240 generated). Then, for both tests, 20 native-language participants were asked to listen to the samples, randomly mixed, and rate them. To evaluate the overall quality, we gathered a \textit{mean opinion score} (MOS) based on a 5-point scale, where 1 stands for ``very poor'' and 5 for ``excellent''. The question asked was: ``How do you rate the overall quality of this sample?''. For accent assessment, we conducted an \textit{accent comparative mean opinion score} (ACMOS) by asking all the participants to rate the samples based on how convincing the accent was perceived to be on a 5-point scale, where 1 stood for ``strong foreign accent'' and 5 for ``native''. The question asked was: ``Does the voice sound like that of a native speaker?''. For this test, Proposed-F was set as the baseline. A negative $\Delta$-ACMOS indicates that a system performs worse than the baseline, zero either refers to the baseline itself or indicates the same performance, and a positive value indicates an improvement over the baseline. The intra-lingual English-English scenario was not considered as the accent is consistent and correct after conversion.

\begin{table*}[t]
\begin{center}
    \resizebox{\textwidth}{!}{
        \begin{tabular}{l|ccccc|ccccc}
            \toprule
            \multicolumn{1}{l|}{} & \multicolumn{5}{c|}{\bf English} & \multicolumn{5}{c|}{\bf French}  \\
            \midrule
            \multicolumn{1}{l|}{} & \bf WER & \bf PER & \bf SSIM & \bf MOS \
             & \bf \(\Delta\)-ACMOS & \bf WER & \bf PER & \bf SSIM  & \bf MOS & \bf \(\Delta\)-ACMOS \\
            \midrule
            Ground truth&   1.25&   0.22&   -&  3.75 ± 0.04&  -& 8.74& 2.70&   -&  3.77 ± 0.05&  -\\
            \midrule
            Discretization &   18.81&   10.28&   90.86&  -& -& 70.00&  60.54&    90.54& -& -\\
            Perturbation &   2.82&   0.88&   \textbf{92.42}&  -& -& 27.17&  12.08&    \textbf{90.84}&  -& -\\
            Utterance std &   4.76&   2.56&   88.62&  -& -& 15.75&  5.58&    84.03& -& -\\
            \midrule
            Soft-VC&   \textbf{2.63}&  \textbf{0.73}&   91.12&  3.88 ± 0.04&  -& 19.13& 6.91&   89.16&  3.73 ± 0.05&  -0.13 \\
            kNN-VC&   2.75&   0.99&   84.70&  3.94 ± 0.04&  -& 16.55& 6.24&   81.26&  3.82 ± 0.05&  -0.10\\
            \midrule
            Proposed&   3.01&   1.09&   90.22&  3.94 ± 0.05&  -& 14.16& 5.00&   88.67&  \textbf{3.85 ± 0.05}&  -0.09\\
            Proposed-F&   2.85&   0.88&   89.70&  \textbf{3.96 ± 0.04}&  -& \textbf{12.43}& \textbf{4.02}&   88.25&  3.83 ± 0.04&  0.0\\
            \midrule
            \midrule
            \multicolumn{1}{l|}{} & \multicolumn{5}{c|}{\bf Spanish} & \multicolumn{5}{c|}{\bf Mandarin Chinese}  \\
            \midrule
            Ground truth&  6.31& 1.52& -& 3.95 ± 0.04 & - & 7.72& 0.43& -& 4.07 ± 0.08& -\\
            \midrule
            Discretization &   75.04& 66.74&   \textbf{92.33}& -& -&  90.01&  77.35&    \textbf{87.26}& -& - \\
            Perturbation &  17.95& 6.52&   89.64&  -& -& 47.73&  20.03&    87.14 & - & - \\
            Utterance std &  12.51& 3.66&   85.43& -& -& 28.34&  5.47&    82.27& -& - \\
            \midrule
            Soft-VC&   14.78& 5.02& 89.17& 3.80 ± 0.04 & -0.08 & 51.82& 21.04& 86.15& 3.01 ± 0.08& -1.25\\
            kNN-VC&   11.91& 3.47& 80.31& 3.83 ± 0.04 & -0.06 & 35.24& 9.78& 77.50& 3.25 ± 0.08& -0.47\\
            \midrule
            Proposed&   9.78& 2.56& 88.13& \textbf{3.87 ± 0.03} & -0.05 & 19.95& 3.64& 84.78& 3.67 ± 0.08& -0.03\\
            Proposed-F&   \textbf{8.61}& \textbf{2.21}& 86.53& 3.86 ± 0.04 & 0.0 & \textbf{18.39}& \textbf{3.33}& 83.02& \textbf{3.73 ± 0.08}& 0.0\\
            \bottomrule
        \end{tabular}
   }
   \caption{Objective and subjective evaluation. (middle) Results (\%) in terms of intelligibility (W/PER \(\downarrow\)) and speaker similarity (SSMI \(\uparrow\)) considering three speaker and content disentanglement strategies: discretization, perturbation and per-utterance standardization. (bottom) W/PER, SSIM, overall quality (MOS \(\uparrow\)) with 95\% confidence intervals, and the delta of the Accent Comparative MOS ($\Delta$-ACMOS) for the proposed approaches and the baseline systems.}
    \label{tab:objective-and-subjective-evalution}
\end{center}
\end{table*}

\subsection{Results}
\label{subsec:results}
In this section, we demonstrate the effectiveness of our proposed approaches evaluating both intra- and cross-lingual scenarios. Additionally, we show that our systems are also capable of preserving the emotions of the source speakers.

\subsubsection{Speaker and Content Disentanglement}
\label{subsubsec:speaker-and-content-disentanglement} 
First, we show that any disentanglement step applied to the SSL representations, while effective in intra-lingual scenario, results in a significant performance loss in cross-lingual tasks. We compared the results obtained by directly using the output of the content encoder against some disentanglement strategies applied on top of the output of the 15th layer of WavLM-Large: discretization, perturbation, and per-utterance standardization. We omitted the soft strategy as it is included in Soft-VC. 

For discretization, we followed the procedure outlined in~\citet{softvc}; for perturbation, we followed the procedure presented in~\citet{nansy}; and for per-utterance standardization, we followed the procedure described in~\citet{per-utt-std}. Table~\ref{tab:objective-and-subjective-evalution} (middle) reports intelligibility in terms of WER and PER, and speaker similarity (SSIM) of each model. Compared to the other methods, the proposed systems significantly enhance the intelligibility across all languages in cross-lingual scenarios while also maintaining very good performance for English. Particularly notable is the significant improvement for Mandarin Chinese, with the WER decreasing from around 30\% to 80\%, and the PER decreasing from approximately 34\% to 95\%. Unlike perturbation, which fails to maintain the same performance observed for English, the overall effectiveness of the per-utterance standardization is significant. However, it does not adequately capture the target speaker's identity like the alternative approaches do. On the other hand, the speaker identity of our systems is comparable to that of other techniques, indicating that in an A2O VC setup, the disentanglement steps are not strictly required.

\subsubsection{Objective and Subjective Evaluation}
\label{subsubsec:objective-and-subjective-evaluation} 
After the initial assessment, we evaluated the quality of the two proposed methods by comparing our outcomes with those of the baselines. Table~\ref{tab:objective-and-subjective-evalution} (bottom) reports WER/PER and SSIM, along with MOS and ACMOS for each model. As observed previously, intelligibility remains clearly superior for the proposed systems for all the cross-lingual tasks and is comparable to that of the baselines for English. In addition, the MOS scores show that both proposed approaches outperform the baselines in terms of overall quality. While kNN-VC shows comparable MOS values, it fails to maintain good speaker similarity. Interestingly, the MOS values of the ground truth are lower than those of all models for English and French. We hypothesize that VC systems can act as denoisers over noisy inputs and that most likely influenced listeners' assignment of MOS scores. As for the accent assessment, Soft-VC and kNN-VC consistently exhibit negative delta values for the ACMOS, meaning that the generated samples are perceived as ``less native-sounding'' across all languages. In conclusion, while both proposed approaches outperform the baselines, Proposed-F leads to better intelligibility, higher MOS, and is also data efficient.

\subsubsection{Ablation: Emotions Maintenance}
\label{subsubsec:emotions-maintenance} 
\begin{table*}[th]
\begin{center}
    \resizebox{\textwidth}{!}{
        \begin{tabular}{l|cccccccc}
            \toprule
            \multicolumn{1}{l|}{} & \bf WER & \bf PER & \bf SSIM & \bf MOS & \(\Delta\)-ECMOS Angry & \(\Delta\)-ECMOS Happy & \(\Delta\)-ECMOS Sad & \(\Delta\)-ECMOS Surprise \\
            \midrule
            Ground truth&   10.08& 4.89& -& 3.95 ± 0.04 & - & -& -& - \\
            \midrule
            Proposed&   11.01& 5.53& 89.54& 3.90 ± 0.05 & +0.39& +0.63& +0.41& +0.92 \\
            Proposed-F&   11.55& 5.80& 88.37& 3.80 ± 0.04 & +0.86 & +0.74& +0.24& +0.80 \\
            \bottomrule
        \end{tabular}
   }
   \caption{Results (\%) measuring intelligibility (W/PER \(\downarrow\)), speaker similarity (SSMI \(\uparrow\)), overall quality (MOS \(\uparrow\)) with 95\% confidence intervals, and the delta of the Emotion Comparative Mean Opinion Score ($\Delta$-ECMOS) vs neutral samples for the emotional experiment.}
    \label{tab:emotions-manteinance}
\end{center}
\end{table*}
In this section we shows our system’s ability to preserve source emotions during VC. To achieve this, we extended the LJSpeech dataset by incorporating the four emotions in the Emotional Speech Dataset (ESD) \citep{esd}: angry, surprise, sad, and happy. As we exclusively focus on English for this experiment, we converted all emotional utterances from the 10 English speakers of ESD into LJSpeech using our highest quality setup. The resulting corpus, referred to as Emo-LJSpeech, consists of roughly 35h of data, 11h of which are created through VC.

To ensure compatibility with the original sample rate of LJSpeech (22050 Hz), we implemented the approach outlined in Section~\ref{subsec:length-regulator} to address length disparities between source and target features. We set $n$-MELs to 80, $n$-fft to 1024, window length to 1024, and hop length to 256. The acoustic model was trained on LJSpeech with a batch size of 32 using AdamW~\citep{adamw} with an initial learning rate of 0.0001 and a weight decay of 0.01 for 200k steps. Throughout the training, the learning rate was linearly decreased after a warm-up period in which it was linearly increased for the first 4k training steps. The above hyper-parameters are the results of an extensive hyper-parameters search phase conducted using a validation set of 100 utterances from LJSpeech. For cross-lingual fine-tuning, we pre-trained the acoustic model on 13h of data from the English subset of ESD (including neutral data). We then fine-tuned it with 2h of LJSpeech for 15k steps using a batch size of 8. For the vocoder, we used an internal implementation of HiFi-GAN working at 22050 Hz trained on a proprietary multi-speaker dataset.

To assess the quality of the newly generated emotional corpus, we created a test set by randomly selecting, for each of the 4 emotions, 3 utterances from 5 speakers extracted from the ESD English subset, resulting in a total of 60 emotional utterances. We then conducted objective and subjective evaluations by comparing the ground truth recordings with the 60 utterances converted into LJSpeech using the \textit{Proposed} and \textit{Proposed-F} models, trained as explained earlier. In addition to assessing WER/PER, SSIM, and MOS, we also analysed the proposed systems' capacity to preserve source emotions through an \textit{emotion comparative mean opinion score} (ECMOS). The test was conducted over each of the four emotions (referred to as \textit{e}). 20 native participants rated the samples on a 5-point scale, indicating the extent to which the corresponding emotion (e.g., happy) was perceived, with 1 denoting ``totally not \textit{e}'' and 5 representing ``very \textit{e}''. The question was: ``Does the voice sound \textit{e}?''. For this test, alongside the 60 emotional utterances, 60 corresponding neutral-style utterances were also generated using both models. These neutral samples served as reference data. We report the $\Delta$-ECMOS which reflects the extent to which a certain emotion was perceived after VC compared to its neutral counterpart. The results are shown in Table~\ref{tab:emotions-manteinance}. The intelligibility and overall quality are very similar to those of the ground truth recordings. Unlike previous experiments, Proposed outperformed Proposed-F. This could be attributed to the limited amount of data used for the pre-training (13h compared to the 150h used in previous experiments). We defer more detailed analysis of this to future work. However, both systems effectively preserve source emotions, as indicated by consistently positive $\Delta$-ECMOS scores. 

\section{Conclusion}
\label{sec:conclusion}
With this work we have developed an A2O VC system capable of generating human-level native-sounding polyglot voices without relying on multilingual data from the target speaker. We showed how several content and speaker disentanglement strategies applied to SSL features do not perform well in cross-lingual scenarios. To overcome this, we have demonstrated that by directly leveraging SSL features and performing cross-lingual fine-tuning, we are able to enhance the intelligibility and accent of the converted voices. Furthermore, we have illustrated that our models can preserve source emotions throughout the conversion process. Our experiments showcased improvements over strong baseline methods across all the examined metrics. Moreover, MOS tests conducted with multiple native participants revealed that our outcomes closely approximate human recordings in terms of overall quality.  For future research, our aim is to explore the creation of high-quality polyglot voices in an A2A setup to further reduce data requirements.

\section*{Limitations}
To achieve high-quality polyglot voices, we focused on A2O and thus target-specific acoustic model. This approach requires a substantial amount of training data. While fine-tuning has helped reduce the required data, 2h can still be considered a significant amount. We leave to future research the development of approaches aimed at further reducing the amount of data needed to train models for specific target speakers.

We conducted our experiments on a single female speaker dataset (LJSpeech). Having limited testing resources, we decided to prioritize the number of experiments on the polyglot aspect over having multiple speakers. This allowed us to show that our model can work correctly with significantly different languages. 

Based on the experiments conducted, the SSIM of the proposed methods is slightly lower, though comparable, to that of other approaches. This discrepancy likely stems from our direct use of SSL features without additional refinements. However, sacrificing a small degree of speaker similarity enhances other aspects such as intelligibility and accent, particularly in cross-lingual tasks. We believe that implementing a more efficient disentanglement scheme will enable us to achieve optimal quality across all components. This is a direction we plan to pursue in future research.

\section*{Acknowledgments}
(Portions of) the research in this paper used the ESD Database made available by the HLT lab, National University of Singapore, Singapore.

\bibliography{custom}

\end{document}